# Characteristics Analysis and State Transfer for non-Markovian Open Quantum Systems

Shuang Cong, Longzhen Hu, Fei Yang, Jianxiu Liu

Department of Automation, University of Science and Technology of China, Hefei, 230027, P. R. China

**Abstract**: The weak-coupled two-level open quantum system described by non-Markovian Time-convolution-less master equation is investigated in this paper. The cut-off frequency $\omega_c$, coupling constant $\alpha$ and transition frequency $\omega_0$, which impact on the system's decay rate $\beta(t)$, coherence factor $C$ and purity $P$, are investigated. The appropriate parameters used in system simulation experiments are determined by comparing analysis results of different values of parameters for the effects of system performance. The control laws used to transfer the system states are designed on the basis of the Lyapunov stability theorem. Numerical simulation experiments are implemented under the MATLAB environment. The features of the free evolution trajectory of the non-Markovian systems and the states transfer from a pure state to a desired pure state under the action of the proposed control laws are studied, respectively. By comparing the experimental results, the effectiveness of the proposed quantum Lyapunov control method applied to the state transfer in non-Markovian open quantum systems is verified. Meanwhile, the influences of different control parameters and cut-off frequencies on the system performance are analyzed.

**Key words**: non-Markovian open quantum systems, dynamics analysis, quantum Lyapunov control, state transfer

## 1 Introduction

Quantum systems can be divided into closed quantum systems and open quantum systems, according to whether systems are isolate with the external environment. The former is under conditions of the absolute zero temperature or does not interact with the external environment and its state evolution is unitary. However, the quantum systems cannot meet the ideal conditions in the practical quantum information processing and quantum computing and have the interaction with the external environment, which are treated as open quantum systems [1-3]. The open quantum systems in which the environment memory effect is ignored can be described by Lindblad-type Markovian master equation under Born or Markovian approximation [4, 5]. This model is widely used in many fields of quantum optics [6]. However, in some other cases, such as the initial state of correlation and entanglement, and most of the condensed matter in which the quantum system interacts with a nanostructure environment, there exists a longer environment memory effect which makes the Markovian approximation invalid and the systems show non-Markovianain characteristics existing in spin echo [7], quantum spot [8] and fluorescence systems [9]. Due to the memory effect existing, non-Markovian quantum systems show more complex characteristics and the state manipulations become more difficult. Recently, people have paid a lot of attention to the researches on the physical characteristics and models of non-Markovian systems, such as non-Markovian dynamics [10-12], entanglement dynamics [13-15] and dynamical models [16]. With the development of quantum control theory and quantum information technology, there is a growth interest in the control of non-Markovian open quantum systems. For instance, optimal control [17], optimal feedback control [18] and coherence feedback control [19] have been successfully applied to the decoherence suppression; Optimal control based on GRAPE (Gradient Ascent Pulse Engineering) [20] and Krotov algorithm [21] were used to search for optimal pulses to implement quantum gates. However, the state-transfer control of non-Markovian open quantum systems, which is a significant and challenging research, has not been widely studied, as far as we know, only involving the state-transfer optimization with a perturbative analysis [22] and the population transfer using SCRAP (Stark-Chirped Rapid Adiabatic Passages) technology [23]. In regard to control methods, the Lyapunov-based control method has the advantages of easy design, analytic form laws and non iterative calculations. The control laws can ensure the system is stable at least. Quantum Lyapunov control method has been widely used in the state preparation and manipulation of closed quantum systems, such as superposition states preparation [24], trajectory tracking [25, 26], state-transfer by optimal control [27] and switching control [28] based on the Lyapunov method and the analysis of convergence with different Lyapunov functions [29, 30]. In addition, some applications of quantum Lyapunov control were studied in Markovian systems [31, 32]. However, to our knowledge, there is little research on state-transfer of non-Markovian systems using this control method.

This paper intends to design control laws based on Lyapunov stability theorem, to control state transfer of non-Markovian systems from an initial pure state to a desired pure state. The weak-coupled two-level open quantum system in high temperature environment described by non-Markovian Time-convolution-less master equation is investigated. The cut-off frequency $\omega_c$, coupling constant $\alpha$ and transition frequency $\omega_0$, which impact on the system's decay rate $\beta(t)$, coherence factor $C$ and purity $p$, are investigated in order to provide the parameters with non-Markovian characteristic for numerical simulation. Considering the fact that the free evolution trajectory of the system state will converge towards the steady state due to the continuous heating effect of the high temperature environment, the control laws based on the Lyapunov stability theorem are designed with variable control parameters. By adjusting the parameters, the task of state transfer from a given initial state to a desired target state is achieved within a given tolerant error. Finally, under the MATLAB environment, two groups of numerical simulations are implemented including the free evolution and the state-transfer under Lyapunov control. The features of the free evolution trajectory in the non-Markovian systems are studied. By comparing the experimental results, the effectiveness of the proposed quantum Lyapunov control method applied to state transfer in non-Markovian open quantum systems is verified. Meanwhile, the influences of different control parameters and cut-off frequencies on the system performance are analyzed.

The paper is organized as follows. In section 2, the control system in high temperature is described by non-Markovian Time-convolution-less master equation. In section 3, the system dynamics with respect to decay rate $\beta(t)$, coherence factor $C$ and purity $p$ in terms of cut-off frequency $\omega_c$, coupling constant $\alpha$ and the transition frequency $\omega_0$ are analyzed. In section 4, based on the Lyapunov stability theorem, the control laws are designed. In section 5, numerical simulations are implemented under the MATLAB environment. The numerical experimental results and the system performance effects are analyzed with different control parameters and cut-off frequencies. Section 6 presents the conclusions.

## 2 The description of the system model

In the weak-coupling limit, assume the form of the interaction Hamiltonians between the system and the environment is bilinear, the two-level controlled system model described by non-Markovian Time-convolution-less master equation can be written as follows[18]:

$$\dot{\rho}_s = -\frac{i}{\hbar}[H, \rho_s] + L_t(\rho_s) \tag{1}$$

$$L_t(\rho_s) = \frac{\Delta(t)+\gamma(t)}{2}\left([\sigma_-\rho_s, \sigma_-^\dagger] + [\sigma_-, \rho_s\sigma_-^\dagger]\right) + \frac{\Delta(t)-\gamma(t)}{2}\left([\sigma_+\rho_s, \sigma_+^\dagger] + [\sigma_+, \rho_s\sigma_+^\dagger]\right) \tag{2}$$

where $H = H_0 + \sum_{m=1}^{2} f_m(t)H_m$ is the controlled Hamiltonian, $H_0 = \frac{1}{2}\omega_0\sigma_z$ and $H_m$ are the system and control Hamiltonian, respectively, $\omega_0$ is the transition frequency of the two-level system, and $f_m(t)$ is the modulation by the time-dependent external control field. The control Hamiltonians can be described by $H_1 = \sigma_x$ and $H_2 = \sigma_y$; $\sigma_x$, $\sigma_y$, $\sigma_z$ are the Pauli matrices $\sigma$; $\sigma_\pm = \frac{\sigma_x \pm i\sigma_y}{2}$ are the rising and lowering operator respectively. For simplicity, we assume $\hbar = 1$.

$L_t(\rho_s)$ describes the interaction between the system and the environment. In Ohmic environment, the analytic expression for the dissipation coefficient $\gamma(t)$ appearing in the equation (2), to second order in the coupling constant, is

$$\gamma(t) = \frac{\alpha^2 \omega_0 r^2}{1+r^2}\left\{1 - e^{-r\omega_0 t}\left[\cos(\omega_0 t) + r\sin(\omega_0 t)\right]\right\} \tag{3}$$

and the closed analytic expression for diffusion coefficient $\Delta(t)$ is [33]:

$$\begin{aligned}
\Delta(t) = \alpha^2 \omega_0 \frac{r^2}{1+r^2}\{&\coth(\pi r_0) - \cot(\pi r_c)e^{-\omega_c t}[r\cos(\omega_0 t) - \sin(\omega_0 t)] \\
&+ \frac{1}{\pi r_0}\cos(\omega_0 t)[\bar{F}(-r_c, t) + \bar{F}(r_c, t) - \bar{F}(ir_0, t) - \bar{F}(-ir_0, t)] \\
&- \frac{1}{\pi}\sin(\omega_0 t)[\frac{e^{-v_1 t}}{2r_0(1+r_0^2)}[(r_0 - i)\bar{G}(-r_0, t) + (r_0 + i)\bar{G}(r_0, t)] \\
&+ \frac{1}{2r_c}[\bar{F}(-r_c, t) - \bar{F}(r_c, t)]]\}
\end{aligned} \tag{4}$$

where $\alpha$ is the coupling constant; $r_0 = \omega_0/2\pi kT$, $r_c = \omega_c/2\pi kT$, $r = \omega_c/\omega_0$ in which $kT$ is the environment temperature, $\omega_c$ is the high-frequency cutoff; $\bar{F}(x,t) \equiv {}_2F_1(x, 1, 1+x, e^{-v_1 t})$, $\bar{G}(x,t) \equiv {}_2F_1(2, 1+x, 2+x, e^{-v_1 t})$, ${}_2F_1(a,b,c,z)$ is the Gauss Hypergeometric function[34].

Under the high temperature limit, one has [33]

$$\Delta(t)^{HT} = 2\alpha^2 kT \frac{r^2}{1+r^2}\{1 - e^{-r\omega_0 t}[\cos(\omega_0 t) - \frac{1}{r}\sin(\omega_0 t)]\} \tag{5}$$

One can see from Eqs. (3) and (5) that for high temperature, both $\gamma(t) \approx 0$ and $|\Delta(t)| \gg \gamma(t)$ hold, demonstrating diffusion coefficient $\Delta(t)$ plays a dominant role in non-unitary dynamics of the system. The essential difference between Markovian systems and non-Markovian systems is the existence of the environment memory effect. Define decay rate as $\beta_{1,2}(t) = \frac{\Delta(t) \pm \gamma(t)}{2}$, then the difference is represented by the sign of $\beta_i(t)$, i.e. when $\beta_i(t) \geq 0$, the system mainly presents the Markovian characteristics; when $\beta_i(t) < 0$, the non-Markovian characteristics are mainly showed [35]. In high temperature, one can easily get $\beta_1(t) \approx \beta_2(t) = \frac{\Delta(t)}{2} = \beta(t)$ since $\gamma(t) \approx 0$. Note that, for medium and low temperatures, the approximation conditions in the Gauss Hypergeometric function used to derive the Eq. (5) are not available, and $\gamma(t)$ can no longer be negligible, then $\beta_i(t)$ are related to both $\Delta(t)$ and $\gamma(t)$.

## 3 Characteristics analyses with different parameters

By analyzing the controlled system (1) in the high temperature, we can find out that the cut-off frequency $\omega_c$, coupling constant $\alpha$ and the transition frequency $\omega_0$ play a key role in the dynamics of open quantum systems. Now, we introduce the coherence factor and purity as the index to weigh the influences of parameters on system dynamics.

As we all know that the density matrix $\rho$ of the two-level system can always be written in the Bloch representation $\rho = \frac{I + \mathbf{r} \cdot \boldsymbol{\sigma}}{2}$, where $\mathbf{r} = (x, y, z) = (tr(\rho\sigma_x), tr(\rho\sigma_y), tr(\rho\sigma_z))$ is a three component real vector and $\|\mathbf{r}\| \leq 1$, so that $\rho = \frac{1}{2}\begin{bmatrix} 1+z & x-iy \\ x+iy & 1-z \end{bmatrix}$. Define the coherence factor as $C = \|x - iy\| = \|x + iy\| = \sqrt{x^2 + y^2}$ and the purity as $p = \text{tr}(\rho_s^2)$. The first-order time derivative of $p$ is given by

$$\begin{aligned} \partial p / \partial t &= 2\text{tr}(\rho_s \dot{\rho}_s) \\ &= 2\text{tr}(\rho_s(-i[H, \rho_s] + L_t(\rho_s))) \\ &= 2\text{tr}(\rho_s L_t(\rho_s)) \end{aligned} \quad (6)$$

Inserting (2) into (6), one has

$$\begin{aligned} \partial p / \partial t &= 2\text{tr}(\rho_s L_t(\rho_s)) \\ &\approx -4\beta(t)\text{tr}(XX^+ - XY - YX + YY^+) \\ &= -4\beta(t)\|X - Y^+\|^2 = -4K\beta(t) \end{aligned} \quad (7)$$

where $X = \rho_s \sigma_-$, $Y = \rho_s \sigma_+$ and $K = \|X - Y^+\|^2 \geq 0$.

From Eq. (7), one can see clearly that the sign of $\partial p / \partial t$ is exactly opposite to the sign of $\beta(t)$ which can be positive or negative, showing the non-monotonicity of the state purity for non-Markovian systems; For closed quantum systems, $\beta(t) = 0$ and the state purity is constant; For Markovian systems, $\beta(t)$ is a constant, thus the sign of $\partial p/\partial t$ is fixed indicating the purity changes is monotonous. Therefore, the variation of purity $p$ displays the obvious differences of the dynamics among the closed systems, the Markovian and non-Markovian systems.

### 3.1 The influence of $\omega_c$ on the decay rate $\beta(t)$

Fixing the ambient temperature $kT$ and transition frequency $\omega_0$, the influence of $\omega_c$ on system dynamics is reflected in the parameter $r = \omega_c/\omega_0$ and $\beta_1(t) \approx \beta_2(t) = \dfrac{\Delta(t)}{2} = \beta(t)$ holds in high temperature. By simple computation for Eq. (5), one can obtain

$$\beta(t) = \alpha^2 kT \frac{r^2}{1+r^2} + \alpha^2 kT \frac{r}{\sqrt{1+r^2}} e^{-r\omega_0 t} \sin(\omega_0 t - \arctan r) \tag{8}$$

Eq. (8) indicates $\beta(t)$ is a continuous, the decreasing function with fluctuation, and it will stabilize at a positive value $\beta_M = \beta(t \to \infty) = \alpha^2 kT \dfrac{r^2}{1+r^2}$. The $r$ value decides the amplitude and attenuation speed of the envelope curve $\Gamma(t) = \alpha^2 kT \dfrac{r^2}{1+r^2} + \alpha^2 kT \dfrac{r}{\sqrt{1+r^2}} e^{-r\omega_0 t}$ and the $\beta_M$. Fig.1 shows plots of $\beta(t)$ for $r = 0.05, 0.1, 1$ within 50 a.u.. We can see from Fig. 1 that for $r = 0.05$, $\beta(t)$ decays slowly over time between positive and negative values with a smaller amplitude and reaches the steady value $\beta_M$ at $t \approx 125$ a.u.. For $r = 1$, $\beta(t)$ holds permanent positive values and reaches $\beta_M$ at $t \approx 9.80$ a.u. with a faster decaying speed. Moreover, in high temperature, $\beta_i(t) \geq 0$ holds for all time when $r \approx 0.274$, and the system described by Eq. (1) degenerates to a Markovian system.

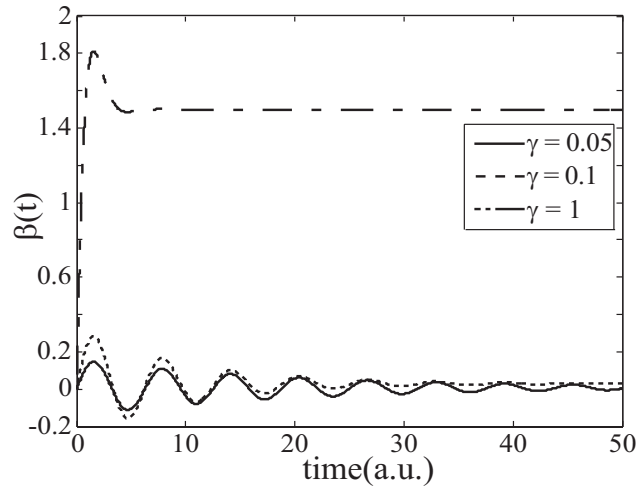

Fig. 1 Plots of $\beta(t)$ with different $r$ values ( $\omega_0 = 1$, $kT = 300\omega_0$, $\alpha = 0.1$ )

From the theoretical analysis in section 2 and Fig. 1, we can see that the system characteristics show a significant difference with different $r$ values. For $r < 0.274$, the non-Markovian and Markovian features are presents by turns, but the former will gradually disappear along with the evolution, thus the system degenerates to a Markovian system and the existing time of non-Markovian properties depends on $r$ value. For $r > 0.274$, $\beta(t)$ holds a faster decaying speed to the steady value and the system mainly presents Markovian features.

### 3.2 The influences of $\omega_c$ on the coherence factor $C$ and purity $P$

In subsection 3.1, we have analyzed the relationship between $r$ and $\beta(t)$. Now a further study on how the cut-off frequency $\omega_c$ affects the state coherence factor $C$ and purity $P$ is done in this subsection. The superposition state $\rho_0 = [1/3\ \sqrt{2}/3; \sqrt{2}/3\ 2/3]$ is chosen as the initial state and Fig. 2 shows the influence of $r = \omega_c/\omega_0$ with the values 0.1 and 1 on the coherence (solid line) and purity (dashed line), respectively. As is shown in Fig. 2, there exists an obvious difference in the system coherence with different $r$. For $r = 0.1$, the coherence gradually decreases with fluctuation; For $r = 1$, the coherence is monotonically decreasing to zero with a faster decaying speed and when $C=0$ holds, the state evolutes to the equilibrium state which indicates the system mainly presents Markovian characteristics. The variation tendency of purity can be explained by Fig. 1, for $\beta(t) > 0$, $P$ falls monotonically and for $\beta(t) < 0$, $P$ raises which abides by Eq. (7), and the purity $P$ changes with fluctuation which can be disappeared along with the evolution and then $\beta(t) > 0$ holds permanently, indicating the disappear of non-Markovian features, and the system degenerates to a Markovian system, then the purity will decrease monotonically. In order to guarantee the appearance of non-Markovian features in the controlled systems, we set $r = 0.05$ in the numerical simulations.

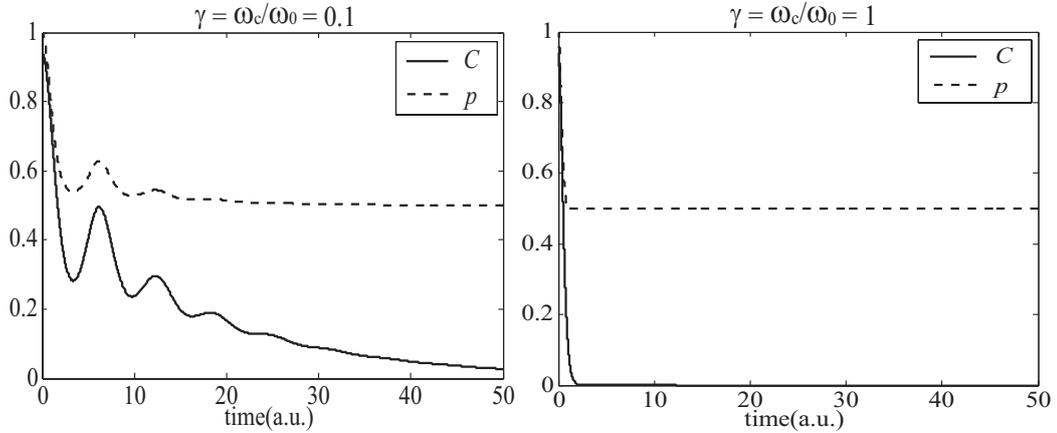

Fig. 2 Plots of coherence and purity with $r = 0.1, 1$

### 3.3 The influences of $\alpha$ on the coherence factor $C$ and purity $P$

The non-Markovian system (1) is obtained based on the second-order perturbation of the coupling item and the limiting condition is the weak-coupling between the system and environment. In this subsection the effect of coupling constant $\alpha^2$ on the system characteristic is studied. The initial state is set as $\rho_0 = [1/3\ \sqrt{2}/3; \sqrt{2}/3\ 2/3]$ chosen in subsection 3.2, Fig. 3 depicts the plots of the coherence (Fig. 3a) and the purity (Fig. 3b) with different $\alpha^2$ values, where the dotted line, solid line, dash-dotted line and dashed line correspond to $\alpha^2 = 0, 0.001, 0.01, 0.05$ respectively.

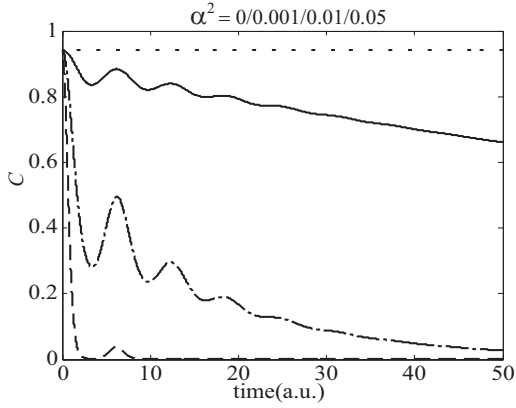 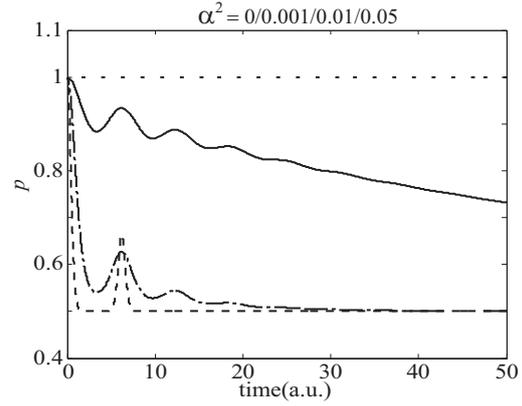

Fig. 3a Curves of the coherence factor $C$      Fig. 3b Curves of the purity $p$

Fig. 3 Curves of the coherence factor $C$ and the purity $p$ with $\alpha^2$ ($\omega_0 = 1$, $kT = 300\omega_0$, $r = 0.1$)

We can see from Fig. 3a that for $\alpha^2 = 0$, the system has no-coupling interaction with environment thus becoming a closed system and the coherence remains unchanged along with the evolution which represents as a horizontal dotted line in Fig. 3a. From Eqs. (3) and (5), one can easily see that $\alpha^2$ is proportional to both the dissipation coefficient $\gamma(t)$ and the diffusion coefficient $\Delta(t)$, thus with the increasing of coupling constant, the decay intensity increases with the same amplitude leading to the coherence curves showing the same frequency with different amplitudes. We can see from Fig. 3b that for $\alpha^2 = 0$, the purity remains a horizontal line with the value 1 due to the pure state being initial state which demonstrates the time evolution operator of the quantum system is unitary. Also, the bigger of the coupling constant $\alpha^2$, the faster of the evolution speed, but the purity does not monotonically decreases along with the evolution which indicates that the memory effects of non-Markovian systems induce a feedback of information from the environment into the system, also, the existence of non-Markovian features and the ability of the information feedback will be strengthened along with the increasing couple constant $\alpha^2$. Note that the coupling constant cannot be a bigger value, like $\alpha^2 = 0.1$, the simulation shows a non-physical behavior, i.e. the state positivity can no longer maintain and the state will jump out of the Bloch sphere for the two-level quantum system, which indicates the chosen coupling constant cannot meet the limiting condition of the controlled system (1). Based on the above analysis, the coupling constant is set as $\alpha = 0.1$ in the numerical simulations.

### 3.4 The influence of $\omega_0$ on the decay rate $\beta(t)$

In subsections 3.1, 3.2 and 3.3, the influences of different environment parameters on the system characteristic are analyzed and we find out that the cut-off frequency $\omega_c$ decides the amplitude and cannot change the frequency of $\beta(t)$. From Eq. (5), one can see that the transition frequency $\omega_0$ influences the decay frequency. In this subsection, we choose different values of $\omega_0$ to observe the influence on $\beta(t)$ and the control performance. In high temperature, the plots of $\beta(t)$ with $\omega_0 = 1, 5, 10$ are shown in Fig. 4 from which one can find out $\omega_0$ only decides the decay frequency which increases with the increasing $\omega_0$ and has no effect on the amplitude of $\beta(t)$. When $\omega_0 = 1$ holds in the numerical simulations, one can see that the state-transfer task may be implemented before the controlled system shows non-Markovian features based on the Lyapunov control laws. With the view of the effect of non-Markovian features on the state-transfer, the transition frequency $\omega_0$ is set as $\omega_0 = 10$ in the numerical simulations.

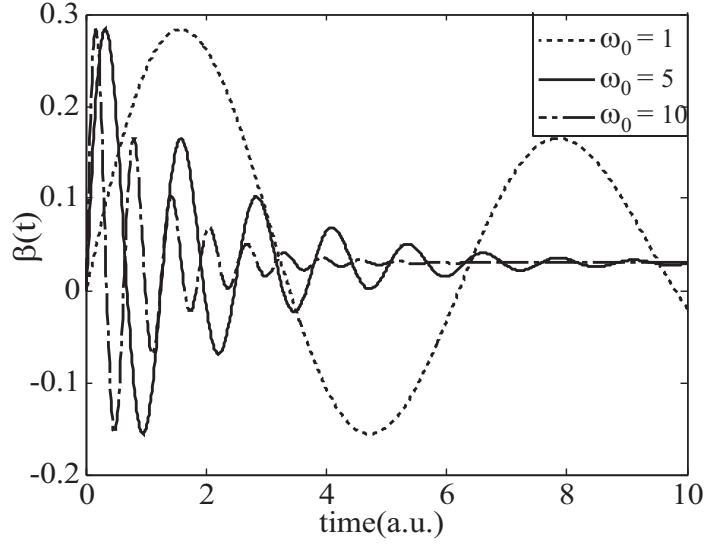

Fig. 4 Plots of $\beta(t)$ with different $\omega_0$ values ($\alpha = 0.1$, $r = 0.1$)

**4 Design of the control laws**

In this paper, we design the control laws based on Lyapunov stability theorem. The basic idea of Lyapunov method is that for $\dot{x} = f(x)$, select a scalar function $V(x)$ with continuous partial derivatives to satisfy the following two conditions: a) $V(x)$ is positive definite, i.e., $V(x) \geq 0$; b) The first order time derivative of the Lyapunov function is negative semi-definite, i.e., $\dot{V}(x) \leq 0$. The designed control laws based on Lyapunov stability theorem can ensure the system is stable at least. The key to design control laws based on Lyapunov method is to select the appropriate Lyapunov function [36].

Based on the state distance, the Lyapunov function is selected as follows:

$$V = \frac{1}{2} tr\left((\rho_s - \rho_f)^2\right) \tag{9}$$

where $\rho_s$ is the controlled state and $\rho_f$ is the desired target state.

Trace distance used to measure the close of two quantum states between $\rho_s$ and $\rho_f$ is $D(\rho_s, \rho_f) = \frac{1}{2} tr|\rho_s - \rho_f| = \frac{1}{2}|r - r_f|$, where $r$ and $r_f$ are the Bloch vectors of $\rho_s$ and $\rho_f$, respectively. Note that the trace distance between two single qubit states is equal to one half the ordinary Euclidean distance between them on the Bloch sphere. The Eq. (9) in the Bloch vector representation can be expressed as: $V = \frac{1}{4}|r - r_f|^2 = D^2(\rho_s, \rho_f)$, therefore we can measure the distance between $\rho_s$ and $\rho_f$ according to the value of the selected Lyapunov function. Define a sufficiently small positive number $\varepsilon$ as the distance error, i.e., as long as $V \leq \varepsilon$ holds during the state transfer, it is believed that the target state is reached under the designed control laws.

The first-order time derivative of $V$ is calculated as

$$\dot{V} = \text{tr}\left(\dot{\rho}_s (\rho_s - \rho_f)\right) \tag{10}$$

Substituting (1) into (10), one has

$$\begin{aligned}\dot{V} &= \text{tr}\left(\dot{\rho}_s (\rho_s - \rho_f)\right) \\ &= \sum_{m=1}^{2} f_m(t) \cdot \text{tr}\left(i \cdot [H_m, \rho_s]\rho_f\right) + \text{tr}\left((L_t(\rho_s) - i[H_0, \rho_s])(\rho_s - \rho_f)\right) \\ &= f_1(t) \cdot T_1 + f_2(t) \cdot T_2 + C\end{aligned} \tag{11}$$

where $T_m = \text{tr}(i \cdot [H_m, \rho_s]\rho_f)$, $m=1,2$ is a real function of $\rho_s$; $f_1$ and $f_2$ are the control laws; $C = \text{tr}((L_t(\rho_s) - i[H_0, \rho_s])(\rho_s - \rho_f))$ is a drift term whose sign can not be determined.

For availability, here we design the control laws ensuring Eq. (11) to satisfy condition (b), i.e. $\dot{V}(x) \leq 0$. The main idea of designing the control laws is that design one control law to offset the influence of the drift term $C$, and design the other control law to ensure $\dot{V}(x) \leq 0$. In the process of designing the control laws, the adjustable threshold variable $\theta$ are introduced, and compared with $T_m$ to determine which control laws to counteract the drift term $C$. The specific design process follows:

(A) In Eq. (11), if $|T_1| > \theta$ holds, then we design the following control law: $f_1 = -C/T_1$, which is to offset $C$, and choose $f_2 = -g_2 \cdot T_2$, $g_2 > 0$. And Eq. (11) becomes $\dot{V} = -g_2 \cdot T_2^2 \leq 0$.

Then the control laws can be written as: $f = \begin{bmatrix} f_1 \\ f_2 \end{bmatrix} = \begin{bmatrix} -C/T_1 \\ -g_2 \cdot T_2 \end{bmatrix}$, where $g_2$ is a positive adjustable control parameter.

(B) In Eq. (11), if $|T_1| < \theta$ and $|T_2| > \theta$ hold, then we design $f_2$ to counteract the drift $C$. Like (A), the designed control laws can be written as: $f = \begin{bmatrix} f_1 \\ f_2 \end{bmatrix} = \begin{bmatrix} -g_1 \cdot T_1 \\ -C/T_2 \end{bmatrix}$, where $g_1$ is a positive control parameter used to adjust the control amplitude, ensuring $\dot{V} = -g_1 \cdot T_1^2 \leq 0$.

(C) In Eq. (11), if $|T_1| < \theta$ and $|T_2| < \theta$ hold, then we calculate the value of Lyapunov function $V$ to estimate the distance between the controlled state and the target state. The control object is deemed to be achieved if the transfer error has reached $\varepsilon$, otherwise we need to reselect the control parameters $g_1$ and $g_2$.

In the following, we clarify the reason why to use the criterion $|T_m| > \theta$ instead of $T_m \neq 0$

to decide which control field to offset $C$. Denoting $\mathbf{r}_f = (x_f, y_f, z_f)$, the expressions of $T_1$ and $T_2$ in the Bloch representation is:

$$T_1 = y_f z - z_f y \tag{12}$$

$$T_2 = z_f x - x_f z \tag{13}$$

From (12) and (13) we can see that: when the controlled state is transferred on the plane $O_1$: $z_f y = y_f z$ or $O_2$: $z_f x = x_f z$, $T_m = 0$ ($m = 1, 2$) hold according to Eqs. (12) and (13), no matter what the value of the control law $f_m$ multiplied with $T_m$ is designed, only $\dot{V} = 0$ holds, showing the value of Lyapunov function $V$ remains unchanged; when the controlled state is transferred on the intersection line $L$ of the planes $O_1$ and $O_2$, which the direction vector is $\mathbf{r}_f$, both $T_1$ and $T_2$ are zero, and $\dot{V} = C$ holds. At this moment, the Lyapunov-based method of designing control laws is unavailable because of the uncertainty of $C$'s sign, and only $V \leq \varepsilon$ holds, is the state-transfer control task from an initial state to a target state considered to achieved, otherwise, we should readjust the control parameters. Then, we can find out that: when $T_m \neq 0$ is used to determine which direction of the control laws to counteract $C$, if and only if $T_m \neq 0$ ($m = 1, 2$) hold, the designed control laws can ensure $\dot{V} < 0$ strictly. Hence, the adjustable threshold variable $\theta$ is introduced in the design process and by judging the size of the two numbers: $T_m$ and $\theta$ to guarantee that the designed control laws can effectively drive the controlled state does not go into the planes $O_1$ and $O_2$, then the first-order time derivative of $V$ can be meet $\dot{V} < 0$ as far as possible, which can ensure that the Lyapunov function $V$ is monotonically decreasing along any dynamical evolution, thus reaching the desired control precision $\varepsilon$. Fig. 5 shows the flow chart of designing control laws according to the above idea in which the execution conditions of the dashed arrow need to meet the following two situations:

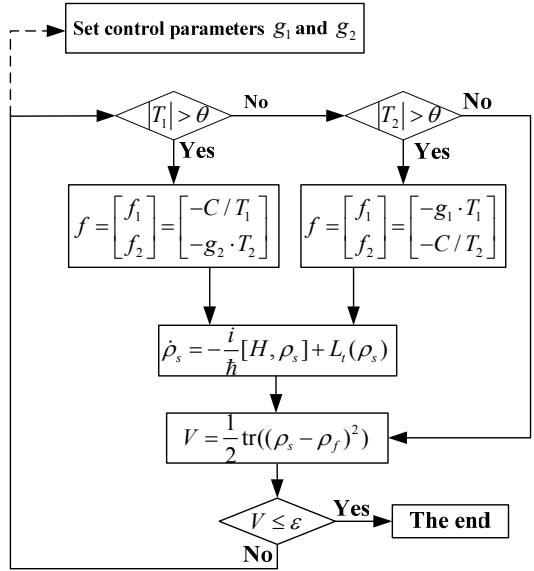

Fig. 5 The flow chart of designing control laws

1) For $|T_1| < \theta$ and $|T_2| < \theta$, when the controlled state is located in the vicinity of the

intersection line $L$, and the transfer error has not reached $\varepsilon$. 2) When the controlled state evolutes in accordance with the designed control laws of (A) and (B) situations, the control task is still not achieved.

**5 Numerical simulations and results analyses**

In this section, numerical simulations are implemented to verify the effectiveness of the designed control laws and the dynamical behavior of non-Markovian systems are analyzed. Numerical simulations are divided into two parts: 1) The free evolution of the uncontrolled system, for which an eigenstate and a superposition state are chosen as the initial state, respectively. The characteristics of non-Markovian systems which are different from that of closed and Markovian quantum systems are illustrated by the analysis of state purity. Meanwhile, this part is considered as a comparative experiment with the second part to verify the effectiveness of the proposed control strategy. 2) The dynamics of the system under Lyapunov control. We have done the following four state-transfer experiments: from an eigenstate to another eigenstate, from an eigenstate to a superposition state, from a superposition state to another superposition state and from a superposition state to an eigenstate. In the four experiments, the dynamics of the controlled system show similar behaviors, thus we take one from a superposition state to another superposition state as an example to analyze and discuss the controlled system characteristics. Based on the content of section 3, the system and environment parameters in the simulation are set as follows: $r = 0.05$, $\omega_0 = 10$, $kT = 30\omega_0$ and $\alpha = 0.1$. Considering the Bloch sphere provides a useful means of visualizing the two-level system state, the evolution trajectories of simulation experiments are represented on Bloch sphere.

**5.1 Free evolutions of the system without the external control fields**

An eigenstate $\rho_{s01}$ and a superposition state $\rho_{s11}$ are chosen to be the initial system states, respectively, they are

$$\rho_{s01} = \begin{bmatrix} 0 & 0 \\ 0 & 1 \end{bmatrix}, \rho_{s11} = \begin{bmatrix} 15/16 & \sqrt{15}/16 \\ \sqrt{15}/16 & 1/16 \end{bmatrix}$$

In order to simulate the limiting state (equilibrium state) of the free evolution, we set a long enough simulation time in the experiments, i.e. the final time is $t_f = 600 a.u.$ with sample time period $\Delta t = 0.1$. The corresponding free evolution trajectories on Bloch sphere are shown in Fig. 6 in which 'o' denotes the initial states, '+' denotes the final point of the state trajectory $\rho_f$. From Fig. 6 one can see that the evolution trajectory is on the $z$ axis of the Bloch sphere when the initial state is the eigenstate $\rho_{s01}$; the evolution trajectory is spiral and arrives at $\rho_f$ progressively while $\rho_{s11}$ is the initial state.

From Fig. 6, one can also see that all the system states eventually terminate on the equilibrium state $\rho_f = diag([0.4917, 0.5083])$ without the external fields whenever the initial states are any kind of the two. Note that, for time $t$ large enough, the coefficients $\gamma(t)$ and $\Delta(t)$ can be approximated by the values $\gamma_M = \gamma(t \to \infty)$ and $\Delta_M = \Delta(t \to \infty)$. From Eqs. (3)

and (5) we have

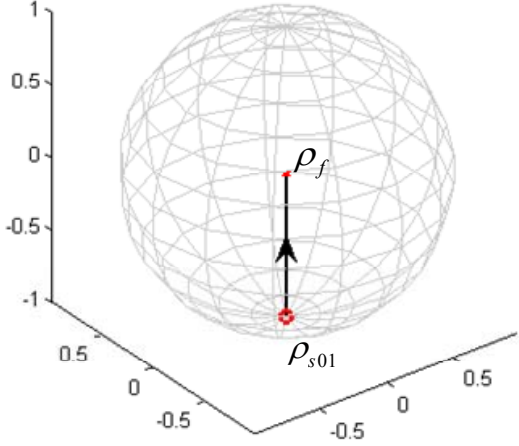 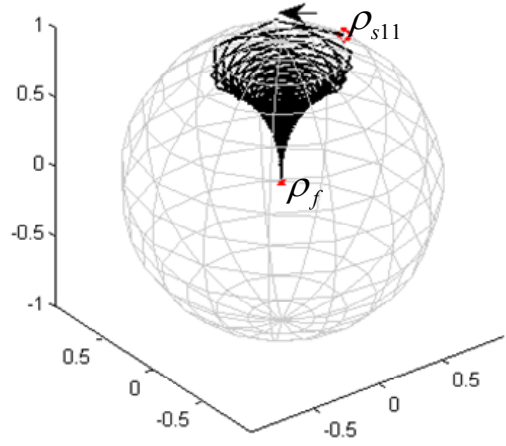

Fig. 6a Initial state is $\rho_{s01}$   Fig. 6b Initial state is $\rho_{s11}$

Fig. 6 The free evolution trajectories with different initial state

$$\gamma_M = \gamma(t \to \infty) = \frac{\alpha^2 \omega_0 r^2}{1+r^2} \tag{14}$$

and

$$\Delta_M = \Delta(t \to \infty) = 2\alpha^2 kT \frac{r^2}{1+r^2} \tag{15}$$

Without the external fields, from $\dot{\rho}_s = 0$, one can get that

$$\begin{bmatrix} -(\gamma_M + \Delta_M z) & -\frac{1}{2}[\Delta_M x + y + i(x - \Delta_M y)] \\ -\frac{1}{2}[\Delta_M x + y - i(x - \Delta_M y)] & (\gamma_M + \Delta_M z) \end{bmatrix} = 0 \tag{16}$$

and it is easy to obtain the coordinate values of the equilibrium state as follows

$$x = 0, y = 0, z = -\frac{\gamma_M}{\Delta_M} = -\frac{\omega_0}{2kT} \tag{17}$$

From Eq. (17) we note that the steady state of the two-level non-Markovian open quantum system is determined by the system transition frequency $\omega_0$ and the environment temperature $kT$, and it has nothing to do with the initial states. With the given parameters, one can easily calculate the density matrix of the steady state $\rho_f = diag([0.4917, 0.5083])$, which coincides with the numerical simulation well.

Take the free evolution of the eigenstate $\rho_{s01}$ as an example to illustrate the features of trajectory and the relationship between $P$ and $\beta(t)$ found out in section 3. Fig. 7 shows the curves of $P$, $\beta(t)$ and $z$ axis figures within 6 a.u. with the initial state $\rho_{s01}$. From Fig. 7, we note that the controlled state on the $z$ axis is not one-way from the South pole of the Bloch sphere to the steady state $\rho_f$, and the values of $z$ axis gradually increase with fluctuation, while the purity $P$ gradually decreases with fluctuation, and both of them have the same turning points, i.e., for $\beta(t) > 0$, $P$ falls and $z$ goes up; for $\beta(t) < 0$, $P$ raises and $z$ grows down.

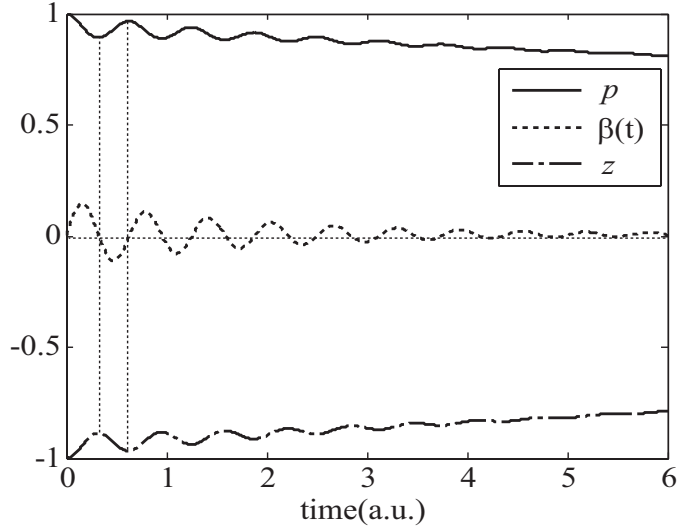

Fig. 7 Curves of $p$, $\beta(t)$ and $z$ axis

**5.2 The states transfer under the Lyapunov control**

In this subsection, the state transfer numerical simulation is implemented from a superposition state to another superposition state and $\Delta t = 5 \times 10^{-4}$. In the experiments, the initial and target states are chosen respectively as

$$\rho_{s11} = \begin{bmatrix} 15/16 & \sqrt{15}/16 \\ \sqrt{15}/16 & 1/16 \end{bmatrix}, \rho_{s12} = \begin{bmatrix} 3/8 & -\sqrt{15}/8 \\ -\sqrt{15}/8 & 5/8 \end{bmatrix}$$

Fig. 8 describes the two state-transfer trajectories on the Bloch sphere under the action of the designed control laws, in which red ' o ' denotes the initial states, '+' denotes the target state and the green ' o ' denotes the final state of the system. Fig. 8a depicts the evolution trajectory from $\rho_{s11}$ to $\rho_{s12}$, and the control parameters are $g_1 = 10$ and $g_2 = 30$. The transfer error between the controlled state and the target state reaches the minimum $\varepsilon = 1.24 \times 10^{-4}$ at 0.6385a.u.. Fig. 8b shows the state transfer from $\rho_{s12}$ to $\rho_{s11}$, whose control parameters are $g_1 = 4$ and $g_2 = 12$, and the transfer error between the controlled state and the target state reaches the minimum $\varepsilon = 1.03 \times 10^{-4}$ at 0.714 a.u.. As can be seen from Fig. 8, under the action of the designed Lyapunov control laws, the state-transfer from different initial states to different target states can be achieved within the desired control performance by adjusting the control parameters. The time-varying control fields $f_1$ and $f_2$ are plotted in Fig. 9 which uses a dual-y coordinate, i.e., the abscissa scale is the same representing simulation time and there are two vertical coordinates including left ordinate describing $f_1$ (solid line) and right ordinate depicting $f_2$ (dashed line). By comparing the evolution trajectories between the free evolution and the state-transfer under the Lyapunov control of the same initial state, one can see that the designed control laws can change the evolution trajectories effectively and drive the controlled state to the target state, and the state-transfer from a pure state to the desired target state of non-Markovian systems is achieved within the given transfer error.

The influence of control parameters on the controlled system performance is illustrated by taking the state-transfer from $\rho_{s12}$ to $\rho_{s11}$ as an example. Fig. 10 shows two trajectories of the state-transfer from $\rho_{s12}$ to $\rho_{s11}$ with different control parameters within the same simulation time 6 a.u., where red ' o ' denotes the initial states, '+' denotes the target state , blue '*'denotes the controlled state at 0.1a.u. and the green ' o ' denotes the final state of the system.

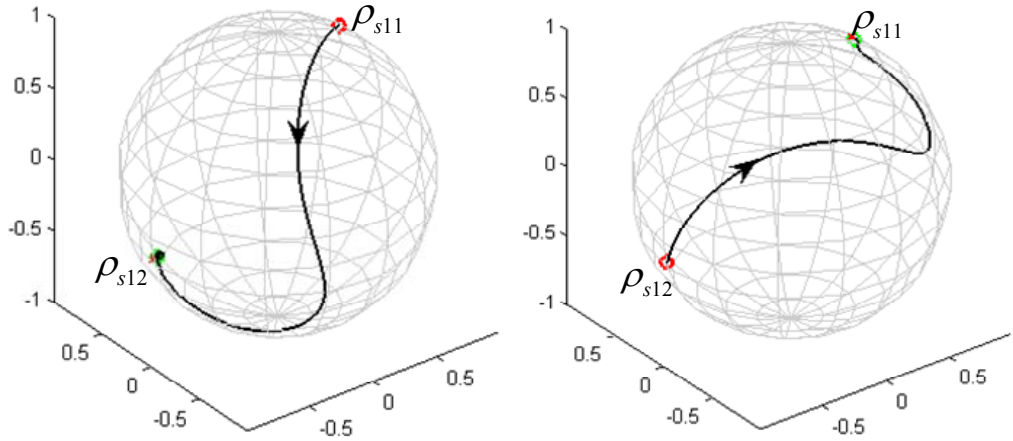

Fig. 8a State-transfer from $\rho_{s11}$ to $\rho_{s12}$    Fig. 8b State-transfer from $\rho_{s12}$ to $\rho_{s11}$

Fig. 8 Two state-transfer trajectories under the designed control laws

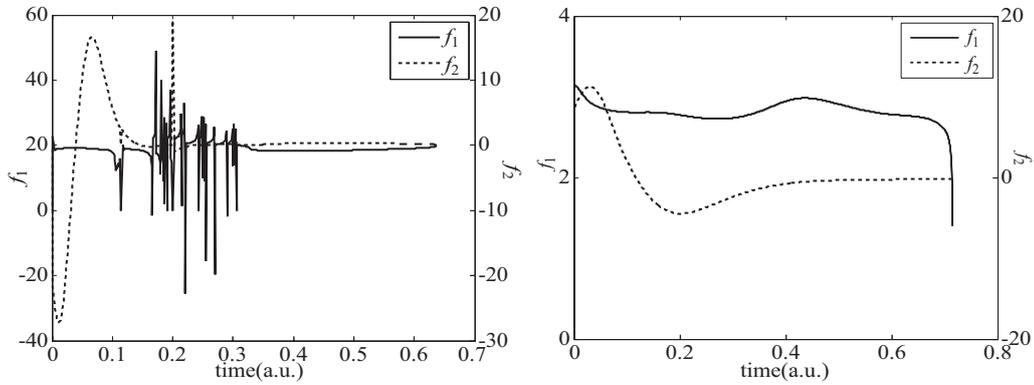

Fig. 9a Control fields from $\rho_{s11}$ to $\rho_{s12}$    Fig. 9b Control fields from $\rho_{s12}$ to $\rho_{s11}$

Fig. 9 Control fields

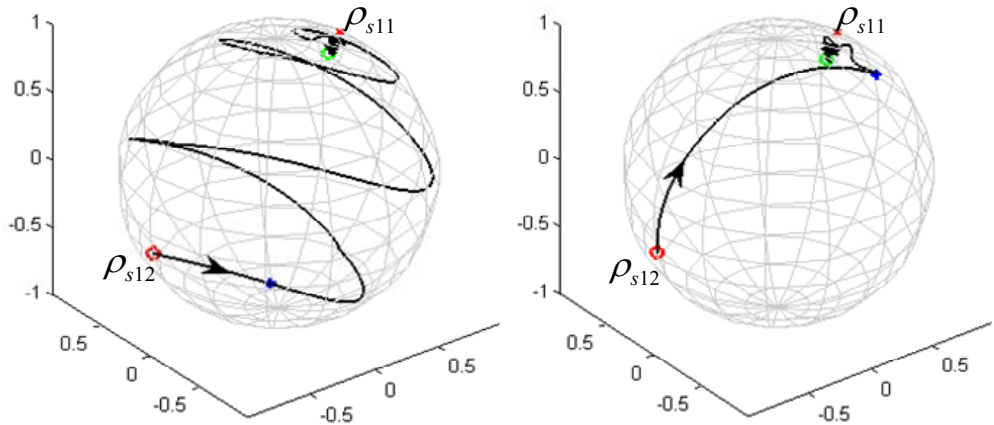

Fig. 10a $g_1 = 4$, $g_2 = 2$    Fig. 10b $g_1 = 4$, $g_2 = 30$

Fig. 10 State-transfer from $\rho_{s12}$ to $\rho_{s11}$ with different control parameters

In Fig. 10a the control parameters are $g_1 = 4$ and $g_2 = 2$ with the minimum transfer error 0.0012; In Fig. 10b the control parameters are $g_1 = 4$ and $g_2 = 30$ with the minimum transfer error 0.0014. As is shown in Fig. 10, in the two situations, the final state has failed to reach at the target state with a less transfer error comparing with Fig. 8b and nor can the transfer error decrease by extending simulation time. Also, we note that control parameters have a great influence on the evolution trajectory at the initial time period, such as $[0,0.1]$, and during the period the controlled state is transferred to the target state at a faster speed along with the increase of $g_2$.

Compare the following situations with different control parameters: Fig. 10a ($g_1 = 4$, $g_2 = 2$), Fig. 10b ($g_1 = 4$, $g_2 = 2$) and Fig. 8b ($g_1 = 4$, $g_2 = 12$) and one can see that for an undersized $g_2$ (Fig. 10a), the evolution trajectory is spiral and has the longest path; For an oversized $g_2$ (Fig. 10b), the controlled state changes faster with a shortened transfer path, but it failed to achieve the given transfer error. Only Fig. 8b has the proper parameters and reaches the performance index $\varepsilon$. Through many experiments, we also find out there exists many groups of control parameters $g_1$ and $g_2$ which can drive the system to reach $\varepsilon$, meanwhile, the groups of control parameter leading to excessive control are also not unique. The optimization of parameters $g_1$ and $g_2$ will be studied in the separate paper.

The effects of $r = \omega_c/\omega_0$ on the controlled system performance are also studied taking the state-transfer from $\rho_{s12}$ to $\rho_{s11}$ with $r = 0.05$ as a contrast. For $r = 0.01$, the control parameters are modulated as $g_1 = 4$ and $g_2 = 10$, we can find out that the control task can be implemented with a less transfer error $\varepsilon = 2 \times 10^{-5}$ at a shorter simulation time of 0.512 a.u.. For $r = 0.1$, a preferred group of control parameters is $g_1 = 4$ and $g_2 = 8$, which can drive the controlled state to reach the minimum transfer error $\varepsilon = 5.1 \times 10^{-4}$ at 0.675 a.u., indicating the degree of the closeness to the target state decreased obviously. For $r = 1$, the controlled system mainly presents Markovian features, by increasing the control action and modifying the control parameters as $g_1 = 4$ and $g_2 = 80$, the obtained transfer error is only up to $10^{-2}$ order and the approximation to the target reduces greatly.

## 6 Conclusions

In this paper, the application of quantum Lyapunov control on state-transfer from a pure state to another pure state of non-Markovian systems has been investigated. The control system in high temperature is described by non-Markovianain Time-convolution-less master equation. The cut-off frequency $\omega_c$, coupling constant $\alpha$ and transition frequency $\omega_0$, which impact on the system's decay rate $\beta(t)$, coherence factor $C$ and purity $p$, have been studied so as to determine the appropriate parameters used in simulation experiments. The control laws have been designed based on the Lyapunov stability theorem. Numerical simulations have been implemented under the MATLAB environment to study the free evolution and the state-transfer from a pure state to a desired pure state under the action of the proposed control laws, respectively. The comparing the experimental results have demonstrated that the free evolution trajectory characteristics of system caused by the change of non-monotonic purity. The designed control laws can effectively achieve the state transfer of a Non-Markovian system from a given initial state to the desired target state. Meanwhile, the control parameters and cut-off frequencies have a significant influence on the controlled system performance.

**Acknowledgements**

This work was supported in part by the National Key Basic Research Program under Grant No. 2011CBA00200, and the National Science Foundation of China under Grant No. 61074050.